\documentclass[aps,prb,twocolumn,reprint,nofootinbib,superscriptaddress,showkeys]{revtex4-2}

\usepackage{color}
\usepackage{graphicx}
\usepackage{dcolumn}
\usepackage{bm}
\usepackage{commath}
\usepackage{amsmath}
\usepackage{upgreek}
\usepackage{siunitx}
\usepackage{physics}
\usepackage[mathlines]{lineno}
\usepackage[english]{babel}
\usepackage[latin1]{inputenc}
\usepackage[bottom=2cm,top=2cm, left=1.5cm, right=1.5cm]{geometry}
\usepackage{booktabs}
\usepackage[unicode=true, bookmarks=true, bookmarksnumbered=false, bookmarksopen=false, breaklinks=false, pdfborder={0 0 1}, backref=false, colorlinks=false]{hyperref}
\usepackage[normalem]{ulem}
\hypersetup{pdftitle={Kondo coherence versus superradiance in THz radiation-driven heavy-fermion system}, pdfauthor={Chia-Jung Yang, Michael Woerner, Oliver Stockert, Hilbert v. Loehneysen, Johann Kroha, Shovon Pal, and Manfred Fiebig}}

\begin{document}

\title{Kondo coherence versus superradiance in THz radiation-driven heavy-fermion systems}

\author{Chia-Jung Yang}
\affiliation{Department of Materials, ETH Zurich, 8093 Zurich, Switzerland}

\author{Michael Woerner}
\affiliation{Max-Born-Institute for Nonlinear Optics and Short Pulse Spectroscopy, 12489 Berlin, Germany}

\author{Oliver Stockert}
\affiliation{Max Planck Institute for Chemical Physics of Solids, 01187 Dresden, Germany}

\author{\hbox{Hilbert v. L\"{o}hneysen}}
\affiliation{Institut f\"{u}r Quantenmaterialien und -technologien and Physikalisches Institut, KIT, 76021 Karlsruhe, Germany}

\author{Johann Kroha}
\email{jkroha@uni-bonn.de}
\affiliation{Physikalisches Institut and Bethe Center for Theoretical Physics, University of Bonn, 53115 Bonn, Germany}
\affiliation{\hbox{School of Physics and Astronomy, University of St.\,Andrews, North Haugh, St.\,Andrews, KY16 9SS, United Kingdom}}

\author{Manfred Fiebig}
\email{manfred.fiebig@mat.ethz.ch}
\affiliation{Department of Materials, ETH Zurich, 8093 Zurich, Switzerland}

\author{Shovon Pal}
\email{shovon.pal@niser.ac.in}
\affiliation{School of Physical Sciences, National Institute of Science Education and Research, An OCC of HBNI, Jatni, 752 050 Odisha, India}

\date{\today}

\begin{abstract}
In strongly correlated systems such as heavy-fermion materials, the coherent superposition of localized and mobile spin states leads to the formation of Kondo resonant states, which on a dense, periodic array of Kondo ions develop lattice coherence. Characteristically, these quantum-coherent superposition states respond to a terahertz (THz) excitation by a delayed THz pulse on the scale of the material's Kondo energy scale and, hence, \textit{independent} of the pump-light intensity. However, delayed response is also typical for superradiance in an ensemble of excited atoms. In this case, quantum coherence is established by the coupling to an external, electromagnetic mode and, hence, \textit{dependent} on the pump-light intensity. In the present work, we investigate the physical origin of the delayed pulse, i.e., inherent, correlation-induced versus light-induced coherence, in the prototypical heavy-fermion compound CeCu$_{5.9}$Au$_{0.1}$. We study the delay, duration and amplitude of the THz pulse at various temperatures in dependence on the electric-field strength of the incident THz excitation, ranging from 0.3 to 15.2\,kV/cm. We observe a robust delayed response at approximately 6\,ps with an amplitude proportional to the amplitude of the incident THz wave. This is consistent with theoretical expectation for the Kondo-like coherence and thus provides compelling evidence for the dominance of condensed-matter versus optical coherence in the heavy-fermion compound.
\end{abstract}

\maketitle

\section{Introduction}
Heavy-fermion (HF) materials are a class of strongly-correlated, rare-earth or actinide intermetallic compounds that exhibit exotic properties, such as non-Fermi-liquid behavior and unconventional superconductivity~\cite{lohneysen2007fermi,steglich1979superconductivity,coleman2007electronedge,wirth2016exploring}. One of the underlying interactions is due to the Kondo effect~\cite{hewson1993book}, an intricate correlation between the localized magnetic moments of the rare-earth 4$f$ or actinide 5$f$ orbitals and the itinerant conduction electrons. At low temperatures this correlation leads to the screening of the local moments and the formation of a spectral resonance peak of width proportional to the Kondo temperature $T_{\text{K}}$ near the Fermi energy $\varepsilon_F$. When the Kondo effect arises in a lattice of localized moments, these local resonances merge into a narrow, hybridized band of width $\sim k_{\text{B}}T_{\text{K}}$ that crosses $\varepsilon_{\text{F}}$. It indicates the existence of heavy fermionic quasiparticles of lifetime $\tau_{\text{K}} = \hbar/k_{\text{B}} T_{\text{K}}$, where $\hbar$ and $k_{\text{B}}$ are the reduced Planck and the Boltzmann constant, respectively~\cite{coleman2007electronedge,coleman2015heavy,ernst2011emerging,paschen2021quantum}.

Recently, dynamic studies on HF materials by THz time-domain spectroscopy (THz-TDS) revealed that heavy-fermion materials respond to an incident single-cycle THz pulse by the emission of a time-delayed THz echo. Upon exciting HF materials with THz radiation, a fraction of the correlated Kondo states is destroyed~\cite{wetli2018time,pal2019fermi,yang2023terahertz,yang2023critical}. It then takes the Kondo coherence time $\tau_{\text{K}}$ to reconstruct the heavy quasiparticle states upon which the relaxing electrons emit a characteristic, echo-like pulse in the THz range at a delay time $\approx\tau_{\text{K}}$, as suggested by a nonlinear rate-equation model~\cite{wetli2018time}. Here the term `nonlinear' refers to the fact that the system response is not governed by the conventional exponential decay~\cite{wetli2018time,yang2023critical}. The nonlinearity and the resulting echo pulse are a fingerprint of strong correlations and coherence intrinsic to the heavy bands and provides information on the quasiparticle dynamics~\cite{wetli2018time,pal2019fermi,yang2023terahertz}. In particular, in this physical picture the echo-pulse intensity and delay time are measures of the quasiparticle weight and their intrinsic coherence time, respectively, and independent of the intensity of the incident pulse. 

Apart from such a nonlinear mechanism originating from the quantum coherence of electronic many-body states inherent to the material, there are other constructive echo-like interference phenomena occurring between spatially disjunct identical subsystems when excited with an intense electromagnetic radiation. Examples are the beating interference of radiation from multiple, equidistant vibrational excitations of CO$_2$ molecules in a molecular gas~\cite{Woerner_OptLett1989} or from different quantum wells in a multiple quantum well sample~\cite{Luo_PRL2004,Shih2005prb}. In addition, there are effects due to experimental geometries, such as the etalon resonance resulting from multiple reflections from parallel-cut sample surfaces~\cite{Rust_OptEng1994}, or trivial reflections from the optical components in the setup.

None of these apply to the metallic, strongly absorbing HF materials, and all the trivial reflections were identified at different delay times~\cite{wetli2018time}. However, the aforementioned inherent source of quasiparticle coherence is challenged by the phenomenon of superradiance which is also a nonlinear effect that can trigger a time-delayed response pulse. The concept of superradiance was introduced in the seminal work by Dicke~\cite{dicke1954coherence} and its dynamics worked out in detail by Rehler and Eberly~\cite{rehler1971superradiance}. The physical process leading to superradiance can be summarized as follows~\cite{scully2009super}. In an ensemble of $N$ identical, excited atoms (more generally, two-level systems), all $N$ excited states can develop phase-coherent time evolution when resonantly coupled via a photonic mode of the electromagnetic environment, that is, when the single-atom excitation energy $\Delta_0=\hbar\omega$ coincides with the mode frequency $\omega$, and if the $N$ atoms are confined within the coherence volume of the mode, typically given by its wavelength $\lambda$. The ensemble then emits an ultrashort, coherent light burst of total field amplitude $E(t)=N\,E_0(t)$ or total intensity $I(t)=N^2\, |E_0(t)|^2$ where $E_0(t)$ is the amplitude of a single emitter. Since the total radiated energy $W$ is the time ($t$) integral of the total emitted intensity,
\begin{equation}
\begin{split}
N \Delta_0 = W = \int \text{d}t\, I(t) & = N^2 \int \text{d}t\, |E_0(t)|^2 \\ & =: N^2  \overline{|E_0|^2} \tau_{\text{s}},
\label{eq:superradiance}
\end{split}
\end{equation}
the emitted pulse duration or decay time must scale as $\tau_{\text{d}} \approx \tau_0/N$, where $\overline{(\dots)}$ indicates the time average and $\tau_0$ is the single-emitter decay time. Time-reversal symmetry implies that the initial build-up time for coherent evolution within the ensemble of $N$ emitters, that is the delay of the emitted pulse, is also given by $\tau_{\text{s}}$~\cite{masson2022universality}. Superradiant phenomena have been observed in defects in semiconductors~\cite{gaal2006nonlinear}, quantum-dot and molecule assemblies~\cite{scheibner2007superradiance,raino2020superradiant}, trapped atomic gases~\cite{goban2015superradiance}, and even in a dense, two-dimensional electron gas~\cite{vasanelli2016ultra}.

Although the Kondo echo and the superradiant response are described by very similar nonlinear differential equations~\cite{wetli2018time,rehler1971superradiance}, there is a key fundamental difference between the two phenomena. The Kondo echo results from intrinsic, dynamical correlations of the material, so that its manifestation is independent of the intensity of the incident pump light. Hence, time-delayed Kondo-induced response is already expected at low intensity. In contrast, the superradiance effect is due to coherence building up by coupling to a virtual electromagnetic mode of the environment. It therefore depends on the intensity of the incident pump light and, thus, is difficult to observe at low intensity. In summary the interplay of condensed-matter coherence and optical coherence in the heavy-fermion compound will therefore depend on the intensity of its excitation. Understanding the physical origin of the delayed pulse in heavy-fermion systems between intrinsic quasiparticle coherence and cooperative decay of radiatively-coupled objects is therefore crucial for extracting the inherent material properties of the system.

In this work, we present a systematic THz-TDS study of the delayed or echo pulse in the prototypical HF compound CeCu$_{5.9}$Au$_{0.1}$, in dependence on the incident THz pump power. CeCu$_{6-x}$Au$_{x}$ undergoes an antiferromagnetic quantum phase transition at the critical Au substitution of $x=0.1$, with strange-metal behavior in electrical transport and thermodynamic properties~\cite{lohneysen2007fermi,stockert2011unconventional}, and is known from earlier THz-TDS measurements~\cite{wetli2018time,yang2020terahertz} to exhibit a pronounced, temperature-dependent echo pulse. For the measurements we chose the quantum-critical compound CeCu$_{6-x}$Au$_{x}$ with $x=0.1$, because understanding the intrinsic quasiparticle dynamics is most pressing at criticality. In the THz-TDS experiment, varying the field strength of the incident THz radiation affects $N$ in the system. Hence, if the emission time of the delayed echo-like response remains unaffected upon changing the field strength, we can associate inherent Kondo correlations with this response. On the contrary, if the emission time correlates with the incident THz field, superradiance is expected to play a role. Here our experiment shows that across the entire range of pump-beam intensities we explored the Kondo-correlations remain the sole identifiable source of the observed quantum-coherent echo-like response. We explain this result by the  nonlinear destruction of the ground-state Kondo spectral weight due to the optical stimulation~\cite{wetli2018time} which does not occur in superradiant~\cite{gaal2006nonlinear,scheibner2007superradiance,raino2020superradiant,goban2015superradiance,vasanelli2016ultra} or linear-interference systems~\cite{Woerner_OptLett1989,Luo_PRL2004}.

\begin{figure*}[t!]
    \centering
    \includegraphics[width=\linewidth]{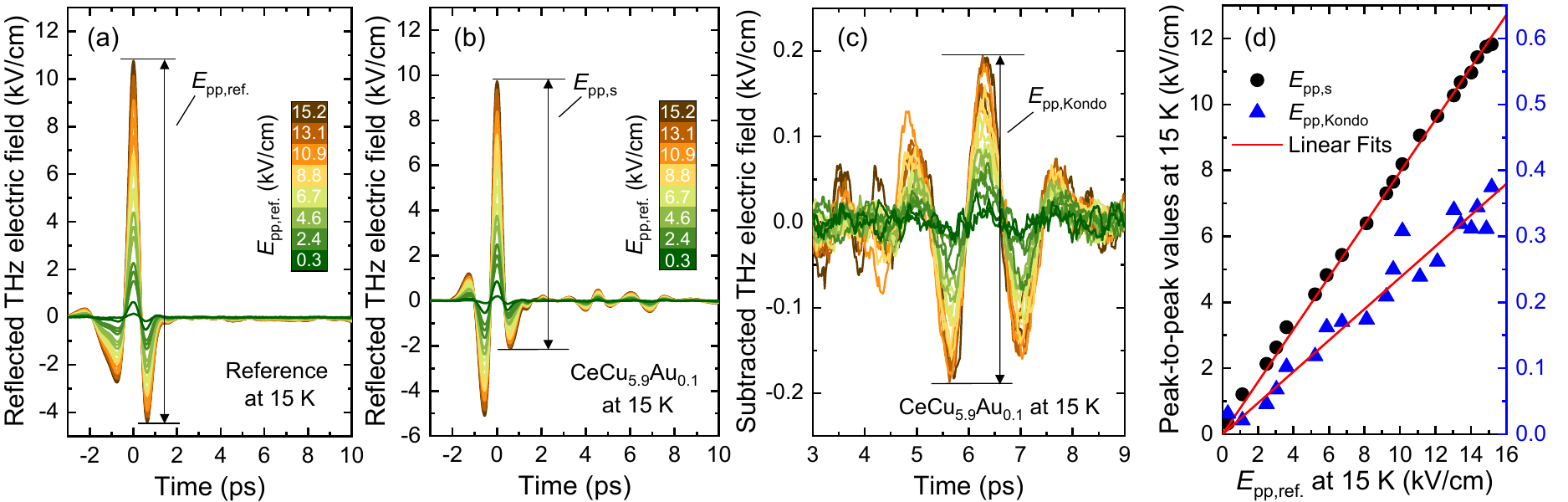}
    \caption{Power-dependence of the THz electric fields reflected from (a) the reference mirror, (b) the quantum-critical compound CeCu$_{5.9}$Au$_{0.1}$ and (c) the background-corrected echo-like Kondo response of CeCu$_{5.9}$Au$_{0.1}$ obtained from (b) at 15\,K. (d) The peak-to-peak values of the instantaneous response ($E_{\text{pp,s}}$) and the Kondo response ($E_{\text{pp,Kondo}}$) of CeCu$_{5.9}$Au$_{0.1}$ obtained from (b) and (c), respectively, as a function of the peak-to-peak values of reference mirror ($E_{\text{pp,ref.}}$) obtained from (a). The fitting curve (red) uses a simple linear function, assuming a zero offset on the $y$ axis.}
    \label{fig:THz_Power_1}
\end{figure*}

\section{Experimental techniques}
\subsection*{Sample preparation} 
The single-crystalline CeCu$_{5.9}$Au$_{0.1}$ samples have been grown by the Czochralski method~\cite{lohneysen1994non,schlager1993magnetic,neubert1997electrical} and are freshly polished at a surface perpendicular to the crystallographic $c$ axis using colloidal silica before the THz-TDS measurements. The surface roughness during sample preparation is within the sub-micrometer range, which is significantly smaller than the wavelength of the incident THz pulse. The samples are mounted in a Janis SVT-400 Helium reservoir cryostat with a controlled temperature environment ranging from 300\,K down to 2\,K.

\subsection*{THz time-domain spectroscopy setup} 
A 1-kHz Ti:Sa regenerative amplifier laser producing 800\,nm pulses (with 120\,fs pulse width and 2.5\,W average power) is utilized for the THz-TDS experiments. The single-cycle, linearly-polarized THz pulses are generated by optical rectification in a 0.5-mm-thick (110)-cut ZnTe generation crystal, using up to 90\% of the amplified laser output. The maximum average power of the incoming infrared beam after the chopper is approximately 442\,mW, at a repetition rate of 500\,Hz. For the field-dependence measurements, we modulate the power of the infrared beam using a variable attenuator which is composed of a half-waveplate to rotate the incoming P-polarized beam and a beamsplitter to selectively transmit its P-polarized component.

The generated THz pulses are guided onto the CeCu$_{5.9}$Au$_{0.1}$ samples. The reflected responses are collinearly focused onto a 0.5-mm-thick (110)-cut ZnTe detection crystal, with an overlap of the sampling pulse using the residual 10\% of the amplified laser output. The THz-induced ellipticity of the sampling pulses is then measured using a quarter-waveplate, a Wollaston prism, and a balanced photodiode. The signals from the balanced photodiode are analyzed with a lock-in amplifier. Under this detection scheme, so-called electro-optic sampling, both amplitude and phase of the THz pulse can be resolved within a single scan. In order to increase the accessible delay time between the THz and the sampling pulses, Fabry-P\'{e}rot resonance from the faces of the detection crystal is suppressed by an additional 2-mm-thick THz-inactive (100)-cut ZnTe single crystal. The additional layer is optically bonded to the back of the detection crystal. All measurements are performed in an inert N$_{2}$-atmosphere.

\begin{figure*}[t!]
    \centering
    \includegraphics[width=0.75\linewidth]{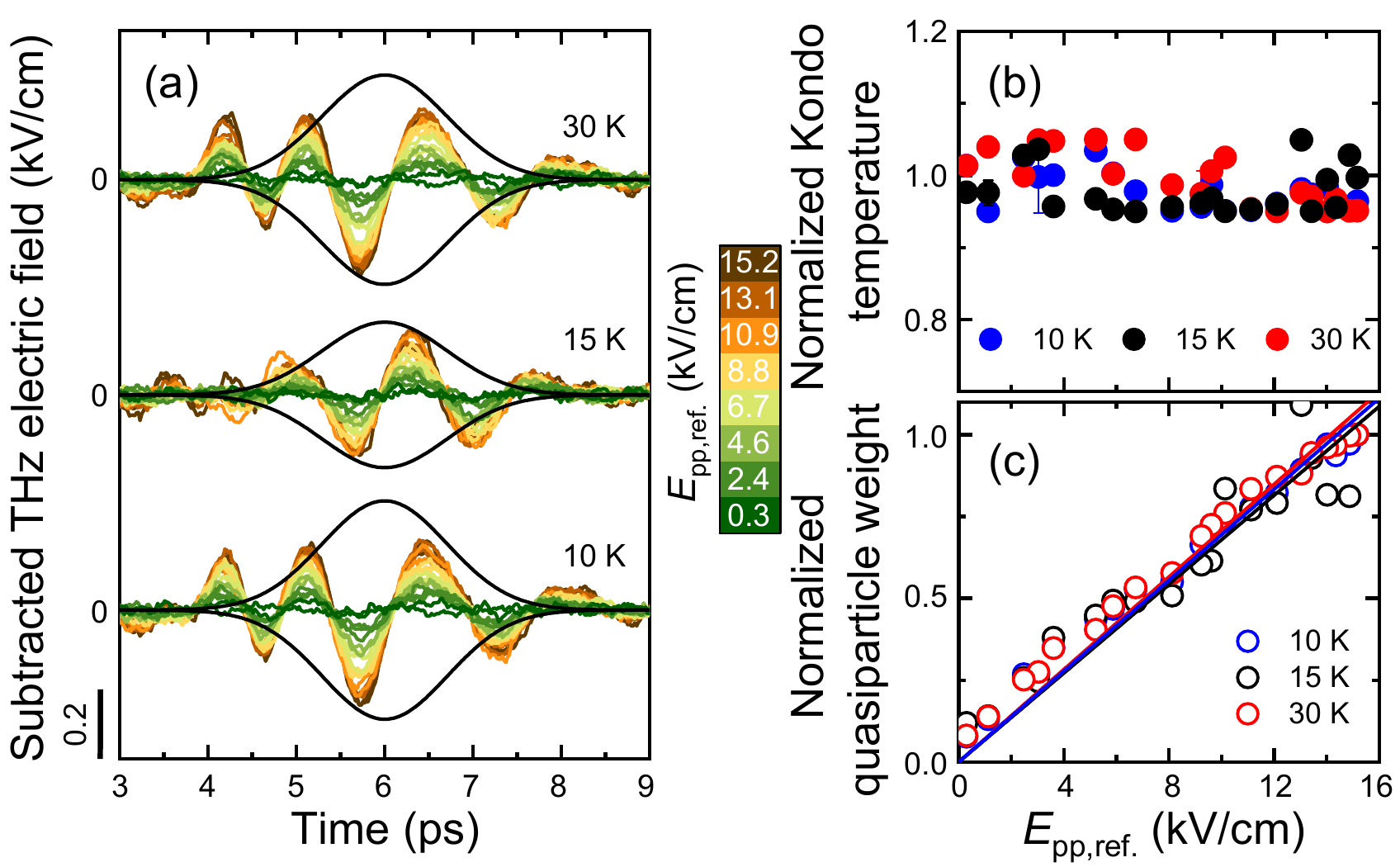}
    \caption{The power-dependence of (a) the background-corrected THz electric fields, (b) the Kondo temperature converted from the fitted delay times of the envelope functions (i.e., the black curves) in (a). (c) The field-dependence of the normalized quasiparticle weight from CeCu$_{5.9}$Au$_{0.1}$ at various temperatures. In (a), the background correction is performed by subtracting the high temperature time trace from the time traces of all temperatures, thereby excluding the high-temperature incoherent signals. The curves are plotted with offsets for better display. In (b), the temperature values are normalized by the literature value of $T_{\text{K}}$. The  fitting curve (red) uses a linear function with zero offset, just as in Fig.~\ref{fig:THz_Power_1}.}
    \label{fig:THz_Power_2}
\end{figure*}

\section{Results and Discussions}
To investigate the behavior of the delayed response in the field-dependent THz-TDS measurements, we first calibrate the THz electric-field strength upon varying the infrared pump power on the THz generation crystal. Figure~\ref{fig:THz_Power_1}(a) shows the THz transients reflected from a Pt mirror, which is used as a reference, at 15\,K. The THz electric fields are converted from the actual read-outs of the lock-in amplifiers. The time transients reflected from the reference mirror have no additional signatures except for the instantaneous response originating from the THz-induced intraband transitions associated with the light conduction electrons~\cite{yang2020terahertz,yang2023terahertz}. We take the peak-to-peak values, denoted by $E_{\mathrm{pp},\mathrm{ref}}$ (i.e., the difference between the maximum and the minimum values of the electric field amplitude) as a measure for the overall field strength of the incident pulse in each of the different THz transients. Within our experimental configurations, the THz field strength lies between 0.3\,kV/cm and 15.2\,kV/cm.

With the knowledge of the field strength of the incident THz pulse, we proceed to examine the field dependence of the delayed response in the quantum-critical compound CeCu$_{5.9}$Au$_{0.1}$. Based on the evolution of quasiparticle spectral weight reported earlier~\cite{wetli2018time}, we select three temperatures for the investigation: 30\,K, 15\,K, and 10\,K. Figure~\ref{fig:THz_Power_1}(b) shows the THz transients reflected from CeCu$_{5.9}$Au$_{0.1}$ at 15\,K, as an illustration. In contrast to the reference mirror (see Fig.~\ref{fig:THz_Power_1}(a)), the THz transients of CeCu$_{5.9}$Au$_{0.1}$ show both the instantaneous and delayed responses. Upon varying the field strength, the peak-to-peak value from both the instantaneous response of CeCu$_{5.9}$Au$_{0.1}$ (denoted as $E_{\text{pp,s}}$, Fig.~\ref{fig:THz_Power_1}(b)) and the delayed Kondo response (denoted as $E_{\text{pp,Kondo}}$, Fig.~\ref{fig:THz_Power_1}(c)) exhibit a linear dependence on the field of the incident THz wave, as shown in Fig.~\ref{fig:THz_Power_1}(d), supported by a linear fit conducted over the entire field range. Such linear dependency implies that the THz-induced intraband and interband transitions are in the non-saturated regime where a large fraction of electrons in the heavy conduction band is still available for further excitations.

We now turn to the key results that concern the field dependence of the delayed response, which may either originate from the THz-induced interband transitions associated with Kondo correlations or from the coherence developed between the optically stimulated atoms associated with superradiance. It is intriguing that in both cases the electric-field amplitude is described by the same rate equation [Eq.~(1) of Ref.~\cite{wetli2018time} for Kondo response and Eq~(4.12) of Ref.~\cite{rehler1971superradiance} for superradiance] and, thus, has the form
\begin{equation}
    E(t) = \frac{E_{0}}{\mathrm{cosh}^{2} \Bigl(2\pi A (\frac{t}{\tau_{\text{d}}}-1) \Bigl)}
    \label{eq_NL-rate-eq_model}
\end{equation}
where $E_{0}$ and $\tau_{\text{d}}$ represent the amplitude and the center (i.e., the delay time) of the field envelope, respectively. Here, the pre-factor $A$ is either equal to unity for the case of Kondo correlations ($A=1$) or proportional to $N$ for the case of superradiance ($A \propto N$)~\cite{rehler1971superradiance}. In the Kondo scenario, Eq.\,(\ref{eq_NL-rate-eq_model}), elucidates the intricate relationship between the Kondo temperature $T_{\text{K}}$ and the delay time (or Kondo coherence time) $\tau_{\text{d}}\equiv\tau_{\text{K}}=h/k_{\text{B}}T_{\text{K}}$~\cite{wetli2018time}. In the superradiant scenario, it describes the sharpening of the delayed-pulse width, $\tau_{\text{d}}=\tau_0/N$, with the excitation density or number of excited atoms $N$, see discussion after Eq.\,(\ref{eq:superradiance}). Note that our observed pulse form $\sim\mathrm{cosh}^{2} \Bigl(2\pi A (\frac{t}{\tau_{\text{d}}}-1) \Bigl)$ corresponds to a Lorentzian line shape in the frequency domain and, thus, indicates that in our experiments there is no significant inhomogeneous broadening, which would otherwise lead to a Gaussian line shape. Figure~\ref{fig:THz_Power_2}(a) shows the delayed, echo-like response emerging within the expected time window (between $+$3.5\,ps to $+$8.5\,ps) after applying a background correction excluding the incoherent signals at high temperature~\cite{wetli2018time}. To extract $E_0$ and $\tau_d$ from the experimental data, we fitted Eq.\,(\ref{eq_NL-rate-eq_model}) to the experimental time-trace envelopes with a fixed value of $A=1$, since using $A$ as an adjustable parameter did not further improve the fit quality. For the sake of presentation, the fitted values of the delay time $\tau_d$ were converted into Kondo temperatures by means of the aforementioned relationship and normalized by the literature value for CeCu$_{5.9}$Au$_{0.1}$ of 8\,K. These normalized values are shown in Fig.~\ref{fig:THz_Power_2}(b) as a function of the field strength $E_{\text{pp,ref}}$ of the incident THz pulse. They vary randomly only within 10\% across the applied THz field range and do not show any tendency of a systematic pump-strength dependence. This essential independence of the delay time on the field strength $E_{\text{pp,ref}}$ verifies that the emitted echo pulses are of Kondo-correlated origin throughout, whereas indications for optically induced superradiance ($\tau_d \propto 1/E_{\text{pp,ref]}}$) are not detected. 

The weight of the emitted THz echos (that is, the square root of the time-integrated modulus squared of the emitted echo THz field, with integration window 3.5\,ps~$\leq t \leq$~8.5\,ps) is shown in Fig.~\ref{fig:THz_Power_2}(c) for selected temperatures. The weights are normalized by the value obtained at the highest THz field strength for each temperature. A linear dependence on the  $E_{\text{pp,ref}}$ over the entire field range is clearly visible. This contrasts the quadratic dependence predicted from superradiance phenomena~\cite{scully2009super,masson2022universality,raino2020superradiant} and, hence, excludes these as the origin of the delayed echo pulse and further corroborates the Kondo scenario. Thus, the weight of the emitted THz echo pulses is a measure of the heavy quasiparticle weight in CeCu$_{5.9}$Au$_{0.1}$, as shown in Fig.~\ref{fig:THz_Power_2}(c).

To better understand why superradiance is not observed in our HF material, let us consider the superradiance and the Kondo delay mechanisms in more detail. In the language of quantum optics, the single emitters of the HF system are the THz-induced excitations of heavy quasiparticles.  These quasiparticle excitations have a single-emitter decay time of roughly equal to the Kondo coherence time, $\tau_{\text{K}}=h/k_{\text{B}} T_{\text{K}}$~\cite{wetli2018time}, and a corresponding coherence length of $\xi_{\mathrm{K}}=v_{\text{F}}\,\tau_{\text{K}}=h\,v_{\mathrm{F}}/k_{\mathrm{B}}T_{\mathrm{K}}$, where $v_{\mathrm{F}}$ is the Fermi speed. The electromagnetic mode in resonance with these excitations (i.e., with frequency $\nu \approx k_{\text{B}} T_{\text{K}}/h$) can establish quantum coherence between these excitations over a distance given by its wavelength, $\lambda=c/\nu = c/v_{\text{F}} \xi_{\text{K}} \gg \xi_{\text{K}}$, with $c$ the speed of light. In principle, such a coherent ensemble of $N$ excitations could emit a superradiant pulse of amplitude proportional to $N^2$ and, thus, of duration $\tau_{\text{K}}/N$ (see discussion of Eq.\,(\ref{eq:superradiance}). However, HF materials are different from semiconductors, atomic gases, and other two-level systems in that the heavy ground-state band exists solely because of the strong Kondo correlations~\cite{hewson1993book}. Upon THz excitation, not only the heavy quasiparticles are excited, but the spectral weight of the heavy band is destructed altogether. This means that the ground-state HF spectral density must be reconstructed during the electronic relaxation process, which occurs on the equilibrium time scale $\tau_{\text{K}}$. Thus, the emission of a short, superradiant pulse of duration $\tau_{\text{K}}/N$ is forbidden.

\section{Conclusions}
In conclusion, our investigation provides fresh insight into the interaction of THz radiation with the quantum-critical compound CeCu$_{5.9}$Au$_{0.1}$. By exploring the excitation-field dependence of the delayed echo-like response, we are able to investigate the mechanism underlying the delayed response by distinguishing the inherent Kondo-induced coherence from the light-induced coherence associated with the superradiance phenomenon. In the entire range of THz electric fields and temperatures applied by us, the robustness of the delay time indicates the dominance of inherent coherence in CeCu$_{5.9}$Au$_{0.1}$. Our result is consistent with the heavy-fermion nonlinear rate-equation model \cite{wetli2018time}, which describes the dynamics of coherent break-up and recovery of Kondo quasiparticle states by THz light. Although we did not observe evidence for superradiance, our findings establish time-domain spectroscopy on the delayed echo pulse as a reliable method to not only investigate quantum materials where the ground-state spectral density is modified and generated by many-body correlation states, such as heavy-fermions, but also investigate other systems with correlated Hubbard-like bands, such as FeSe~\cite{Watson2017prb}, as opposed to semiconductor single-particle bands.

\section*{Acknowledgements}
This work was financially supported by the Swiss National Science Foundation (SNSF) via project Nos.\,200021\_178825 (M.F., C.-J.Y.) and 200021\_219807 (M.F.) and by the Deutsche Forschungsgemeinschaft (DFG) via SFB/TRR 185 (277625399) OSCAR (project C4) and the Cluster of Excellence ML4Q (90534769) (J.K.). S.P. acknowledges the start-up support from DAE through NISER and the project Basic Research in Physical and Multidisciplinary Sciences via RIN4001. In addition, S.P. also acknowledges the support from SERB through SERB-SRG via Project No.\,SRG/2022/000290.

\end{document}